\documentclass[12pt]{iopart}
\usepackage{hyperref,graphicx,latexsym,amssymb,verbatim,revsymb,color}

\renewcommand{\Tr}{\textrm{Tr}} 
\newcommand{\bra}[1]{\langle #1|} 
\newcommand{\ket}[1]{|#1\rangle}
\usepackage{mathpazo}
\usepackage{hyperref}

\begin{document}

\title{The smallest refrigerators can reach maximal efficiency}

\author{Paul Skrzypczyk$^1$, Nicolas Brunner$^1$, Noah Linden$^2$ and Sandu Popescu$^1$} 
\address{$^1$ H. H. Wills Physics Laboratory, University of Bristol, Tyndall Avenue, Bristol, BS8 1TL, United Kingdom \newline $^2$ Department of Mathematics, University of Bristol, University Walk, Bristol BS8 1TW, United Kingdom }

\date{\today} 
\begin{abstract}
We investigate whether size imposes a fundamental constraint on the efficiency of small thermal machines. We analyse in detail a model of a small self-contained refrigerator consisting of three qubits. We show analytically that this system can reach the Carnot efficiency, thus demonstrating that there exists no complementarity between size and efficiency. 
\end{abstract}

\maketitle

\section*{Introduction} 

Quantum thermodynamics, which investigates physical phenomena at the intersection of quantum mechanics and thermodynamics, is by now a well developed field \cite{QT1,QT2,QT3}. In particular, there has been significant interest in quantum heat engines \cite{H1,H2,H3,H4,H5,H6} as well as refrigerators \cite{F1,F2}. Quantum analogues of Carnot engines have been studied extensively \cite{C1,C2,C3,C4} as well as other cycles, such as Otto cycles \cite{O1,O2,O3,O4} and Brownian motors \cite{B1}. There has also been an interest from the perspective of quantum information \cite{I1}. 

Recently, a growing interest has been devoted to small \emph{self-contained} quantum thermal machines \cite{LinPopSkr09,HE,palao,youssef}. By self-contained we mean that no sources of external work or other form of control are allowed; only incoherent interactions with thermal baths at various temperatures. Interestingly it was shown that there exist no fundamental limitation on the size of such quantum thermal machines. Here our main focus will be to determine how efficient such small machines can be; whether size imposes fundamental limitations upon its efficiency.

In a Carnot engine the machine passes adiabatically through \emph{very many} states; indeed through infinitely many of them. For example in the case of an ideal gas contained within a cylinder, a piston is slowly moved, with the volume of the gas changing continuously from its initial to final value. However, for the case of a quantum fridge with only a small number of distinct states it is no longer the case that we can transition through many states, a design constraint which appears at first sight rather drastic. The question therefore arises as to whether this imposes an additional bound which will prevent us from achieving an efficiency equal to, or even close to, the Carnot limit. It is conceivable then that this bound tends towards the Carnot efficiency as the number of states increases towards infinity. One may therefore ask whether or not there exist in nature a complementarity between size and efficiency? Is it the case that, in order to be efficient, thermal machines must be large, having access to many states, or can small machines be efficient as well? 

Remarkably, it turns out that there exists no such complementarity between size and efficiency; machines with only a small number of states can reach the Carnot efficiency. We demonstrate this for the three-qubit model of the smallest self-contained refrigerator recently presented in 
Ref.~\cite{LinPopSkr09}. This model is particularly simple and transparent allowing us to give an exact analytic treatment of the stationary behaviour for all values of the parameters of the model.  This leads to an extremely simple formula for the efficiency; we can then show that this efficiency can be made arbitrarily close to the Carnot limit. In \cite{palao,schultz}, similar conclusions (that one can reach Carnot efficiency) were reached for other (self-contained or work-driven) small thermal machines.

\section*{The model} To start with let us introduce more precisely the model which we will focus on. As stated above, this is a model of small, self contained refrigerators. By small we mean that we consider quantum systems composed of very few states, and by self contained we mean that we consider systems whose internal evolution is governed by a time-independent Hamiltonian and whose supply of free energy comes solely through contact with thermal reservoirs at differing temperatures; therefore no external work is involved. We showed that it is possible to construct refrigerators meeting our requirements, hence demonstrating that there is no fundamental limit on the size of such thermal machines. Initially let us consider 3 non-interacting qubits. The free Hamiltonian for the three particles is given by
\begin{equation}\label{e:Ham}
	H_0 = H_1 + H_2 + H_3 = E_1 \Pi_1 + E_2 \Pi_2 + E_3 \Pi_3
\end{equation}
where $\Pi_i = \ket{1}_i\bra{1}$ is the projector onto the excited state for each particle. We will constrain the energy levels such that $E_2 = E_1 + E_3$ for reasons which will become evident.

We take each qubit to be in contact with a thermal reservoir. The temperature of each reservoir will be taken to be different; we denote the temperatures of the reservoir of qubits 1, 2 and 3 as $T_C$, $T_R$ and $T_H$ respectively, which we will refer to as the ``\emph{cold}'', ``\emph{room}'' and ``\emph{hot}'' reservoirs. To model the process of thermalisation of each qubit by the bath, we take a simple reset model, whereby with probability density $p_i$ per time $\delta t$ each qubit may be reset to a standard thermal state $\tau$ of its bath. Formally this amounts, in time $\delta t$, to the non-unitary process
\begin{equation}\label{e:dissapator}
	\rho \mapsto \sum_i p_i\delta t \tau_i \otimes \Tr_i \rho + (1-p_i\delta t)\rho
\end{equation}
where, taking $k_B = 1$, $\tau_i = e^{-H_i/T_i}/Z \equiv r_i\ket{0}_i\bra{0} + \overline{r_i}\ket{1}_i\bra{1}$ is the Boltzmannian, $Z = \Tr e^{-H/T_i}$ is the partition function and $r_i$ and $\overline{r_i}$ are the probabilities for the $i^{th}$ qubit to be in the ground and excited state respectively, given by
\begin{equation}\label{e:ri}
	r_i = 1/(1+e^{-E_i/T_i}),\quad\quad\quad\quad \overline{r_i} = e^{-E_i/T_i}/(1+e^{-E_i/T_i})
\end{equation}

To turn this system into a refrigerator we introduce the interaction Hamiltonian $H_{int}$,
\begin{equation}
	H_{int} = g(\ket{010}\bra{101} + \ket{101}\bra{010})
\end{equation}
which couples the three particles. Given the imposed constraint, $E_2 = E_1 + E_3$, this Hamiltonian couples only states degenerate in energy. 

Furthermore we consider only the scenario in which this interaction is \emph{weak}, that is we take $g \ll E_i$. In this regime the interaction Hamiltonian does not appreciably alter the energy eigenvalues or eigenvectors of the system, which remain governed by $H_0$. It is therefore justified to define the thermal state of the qubits which depends only upon $H_0$. Note also that in general the addition of $H_{int}$ between the particles requires a modification of the dissipative dynamics, \Eref{e:dissapator}, if it is to remain consistent \cite{T1}. However, we are interested only in the limit where $g$ and $p_i$ vanish such that $g/p_i$ remains constant. Since corrections to the dissipative dynamics are of order $pg$ or higher, in this limit they vanish, and hence \Eref{e:dissapator} remains a consistent dynamics.

In \cite{LinPopSkr09} a detailed analysis of this model was given and it was shown why it behaves as a refrigerator. Here, to briefly understand the basic idea behind its functioning, notice that the Hamiltonian simply interchanges the population of the states $\ket{010}$ and $\ket{101}$. In the absence of any interaction with the environment the transitions in either direction are equiprobable and therefore nothing is achieved. However, by introducing environments at different temperatures, $T_R < T_H$, the interaction can be made `biased' so that the final occupation probabilities of the states relative to their values at thermal equilibrium become significantly altered. In this way a refrigerator can thus be constructed; the end result is that qubit one reaches a stationary temperature lower than that of its environment, $T_1^{S} < T_C$.

Since the qubits interact with an environment the dynamics is described by a master equation. The master equation governing the dynamics of the refrigerator is given by
\begin{equation}
	\frac{\partial \rho}{\partial t} = -i[H_0+H_{int},\rho] + \sum_{i=1}^3 p_i(\tau_i \otimes \Tr_i \rho - \rho).
\end{equation}

We are interested in the stationary (or long term) behaviour of the system, and thus wish to find $\rho^{S}$ satisfying 
\begin{equation}
	0 = -i[H_0+H_{int},\rho^{S}] + \sum_{i=1}^3 p_i(\tau_i \otimes \Tr_i \rho^{S} - \rho^{S}).
\end{equation}
This equation can be solved exactly and analytically. It can be checked straightforwardly that the solution is given by
\begin{eqnarray}
	\rho^{S} &=& \tau_1\tau_2\tau_3 + \gamma\Big(Q_{23} Z_1\tau_2\tau_3 + Q_{13} \tau_1 Z_2 \tau_3 + Q_{12} \tau_1 \tau_2 Z_3\  \nonumber \\
	&& + q_1 \tau_1 Z_{23} + q_2 \tau_2 Z_{13} + q_3 Z_{12}\tau_3 + Z_{123} + \frac{q}{2g}Y_{123}\Big)
\end{eqnarray}
where $Y_{123} = i\ket{101}\bra{010}-i\ket{010}\bra{101}$ and $Z_{123} = \ket{010}\bra{010}-\ket{101}\bra{101}$ are Pauli-like operators, $Z_{12} = \Tr_3 Z_{123}$ and so fourth. Furthermore the parameters $q_i$ and $Q_{jk}$, depending only upon the thermalisation rates $p_m$, are given by
\begin{equation}
	q_i = \frac{p_i}{q-p_i},\quad\quad\quad\quad\quad\quad\quad\quad Q_{jk} = \frac{p_jq_k + p_kq_j}{q-p_j-p_k}, 
\end{equation}
where $q = p_1 + p_2 + p_3$. Finally, the parameter $\gamma$ is given by
\begin{equation}\label{e:gamma}
	\gamma = \frac{-\Delta}{2 + \frac{q^2}{2g^2} + \sum_i q_i + \sum_{jk} Q_{jk}\Omega_{jk}}
\end{equation}
where
\begin{equation}\label{e:delta}
	\Delta = r_1 \overline{r_2} r_3 - \overline{r_1} r_2 \overline{r_3}, \quad\quad\quad\quad \Omega_{jk} = r_j'\overline{r_k}' + \overline{r_j}'r_k.
\end{equation}
Here $r'_i = \overline{r_i}$ for $i=2$, otherwise $r'_i = {r_i}$. The first notable features of the solution is that all single-party and two-party reduced density matrices are diagonal. Second, and of most importance, is the form of the single-party states which is given by
\begin{equation}\label{e:rho i}
	\rho_i^S = \tau_i + \frac{q\gamma}{p_i} Z_i
\end{equation}
thus the occupation probability of the ground state for each qubit is shifted from its value at equilibrium by an amount proportional to the parameter $\gamma$. 

For this model to act as a refrigerator, the stationary temperature of qubit 1 must be colder than its bath temperature, i.e. $T_1^S < T_c$. This happens whenever the occupation probability of the ground state for particle 1 is increased compared to its thermal population. This happens whenever $\gamma > 0$. From \Eref{e:gamma} it can be checked that the denominator is a positive quantity and therefore the sign of $\gamma$ depends only upon the numerator, $-\Delta$. Using the definitions \Eref{e:delta} and \Eref{e:ri} it can be shown that the condition $-\Delta > 0$ is equivalent to
\begin{equation}
	e^{-E_1/T_C}e^{-E_3/T_H} > e^{-E_2/T_R}
\end{equation}
which, upon further manipulation, can be re-expressed as
\begin{equation}\label{e:bound}
	\frac{E_1}{E_3} < \frac{1-\frac{T_R}{T_H}}{\frac{T_R}{T_C}-1}
\end{equation}
This is the fundamental design constraint on our refrigerator; as long as this condition is satisfied our model works as a refrigerator. As the ratio $E_1/E_3$ approaches the above limit, the temperature of the cold qubit approaches from below the temperature of its bath; everything else being held constant, this implies that it will take longer for the refrigerator to draw heat from the cold bath, similarly to what happens to a classical refrigerator as one approaches the reversible limit, as its functioning becomes adiabatically slow. The above fundamental design constraint will play the central role in analysing the efficiency. 

\section*{The quantum efficiency}
To analyse the efficiency of the refrigerator an expression for the amount of heat that the quantum machine is able to exchange with the thermal reservoirs in which it is in contact must be derived. To do this let us consider the change of one of the particles in a small time $\delta t$ induced by the resevoir. From \Eref{e:dissapator} we find that
\begin{eqnarray}
	\delta \rho_i(t) &=& \rho_i(t+\delta t) - \rho_i(t) = p_i \delta t \tau_i + (1- p_i\delta t) \rho_i(t), \nonumber \\
	&=& p_i \delta t (\tau_i - \rho_i(t)).
\end{eqnarray}
To this change of state corresponds a change in energy, $\delta\mathcal{E}_i$, given by 
\begin{equation}
	\delta \mathcal{E}_i = \Tr(H_i \delta \rho_i(t)) = p_i\delta t \Tr(H_i(\tau_i-\rho_i(t))
\end{equation}
thus, taking the limit $\delta t \to 0$ gives the rate of change of energy of the particle due to the interaction with the reservoir
\begin{equation}
	\frac{d\mathcal{E}_i}{dt} = p_i \Tr(H_i(\tau_i-\rho_i(t))
\end{equation}
which in other words it is the amount of energy supplied to the particle from the bath and is therefore the rate of heat flow, which we shall denote $Q_i$.

Using the explicit form previously obtained for $\rho_i$, \Eref{e:rho i} along with the definition of the Hamiltonian \Eref{e:Ham} we find that this can be re-written as
\begin{equation}
	\frac{d\mathcal{E}_i}{dt} = p_i\Tr\Bigg(E_i\Pi_i\bigg(-\frac{q\gamma}{p_i}Z_i\bigg)\Bigg) = (-1)^{i+1} q\gamma E_i,
\end{equation}
where the factor $(-1)^{i+1}$ arises due to the fact that $Z_1 = -Z_2 = Z_3 = Z$, the standard Pauli operator. Thus we see that the rate of heat flow between each bath and particle is given by
\begin{equation}
	Q_C = q\gamma E_1, \quad\quad\quad Q_R = -q\gamma E_2, \quad\quad\quad Q_H = q\gamma E_3,
\end{equation}
and thus the efficiency of our quantum refrigerator is given by
\begin{equation}\label{e:quantum efficiency}
	\eta^Q = \frac{Q_C}{Q_H} = \frac{E_1}{E_3}.
\end{equation}
We arrive at the interesting result that although the individual heat currents have a rather complicated dependence upon all of the parameters in the problem, through $q$ and $\gamma$, the efficiency of the fridge is in fact independent on all parameters except the ratio of energy levels. This result, although at first sight contradictory, is consistent with the results found in \cite{H3} and can be understood qualitatively: It is the interaction Hamiltonian which takes the particles away from their thermal equilibrium states, and since the Hamiltonian only acts on particles 1 and 3 simultaneously its clear that the rates at which they exchange heat with their reservoirs must be proportional to each other -- hence the dependence in each case cancels when looking at the ratio.

Equation \Eref{e:quantum efficiency} however must be taken in conjunction with the basic design constraint, equation \Eref{e:bound}, which then yields an upper bound on the quantum efficiency:
\begin{equation}
	\eta^{Q} < \frac{1-\frac{T_R}{T_H}}{\frac{T_R}{T_C}-1}.
\end{equation}
It is important to note that since the refrigerator works as long as the condition \Eref{e:bound} is satisfied that this is indeed an achievable bound on the efficiency of the refrigerator. In other words, we can get as close as we like to the following quantum efficiency
\begin{equation}\label{e:quantum max}
	\eta^{Q}_{max} = \frac{1-\frac{T_R}{T_H}}{\frac{T_R}{T_C}-1}.
\end{equation}

\section*{The Carnot efficiency}
In order to see the significance of the above derived maximum quantum efficiency for our particular model, we need to compare it with the Carnot efficiency derived from a `standard' model. 

\begin{figure}[h] 
	\includegraphics[width=0.7\columnwidth]{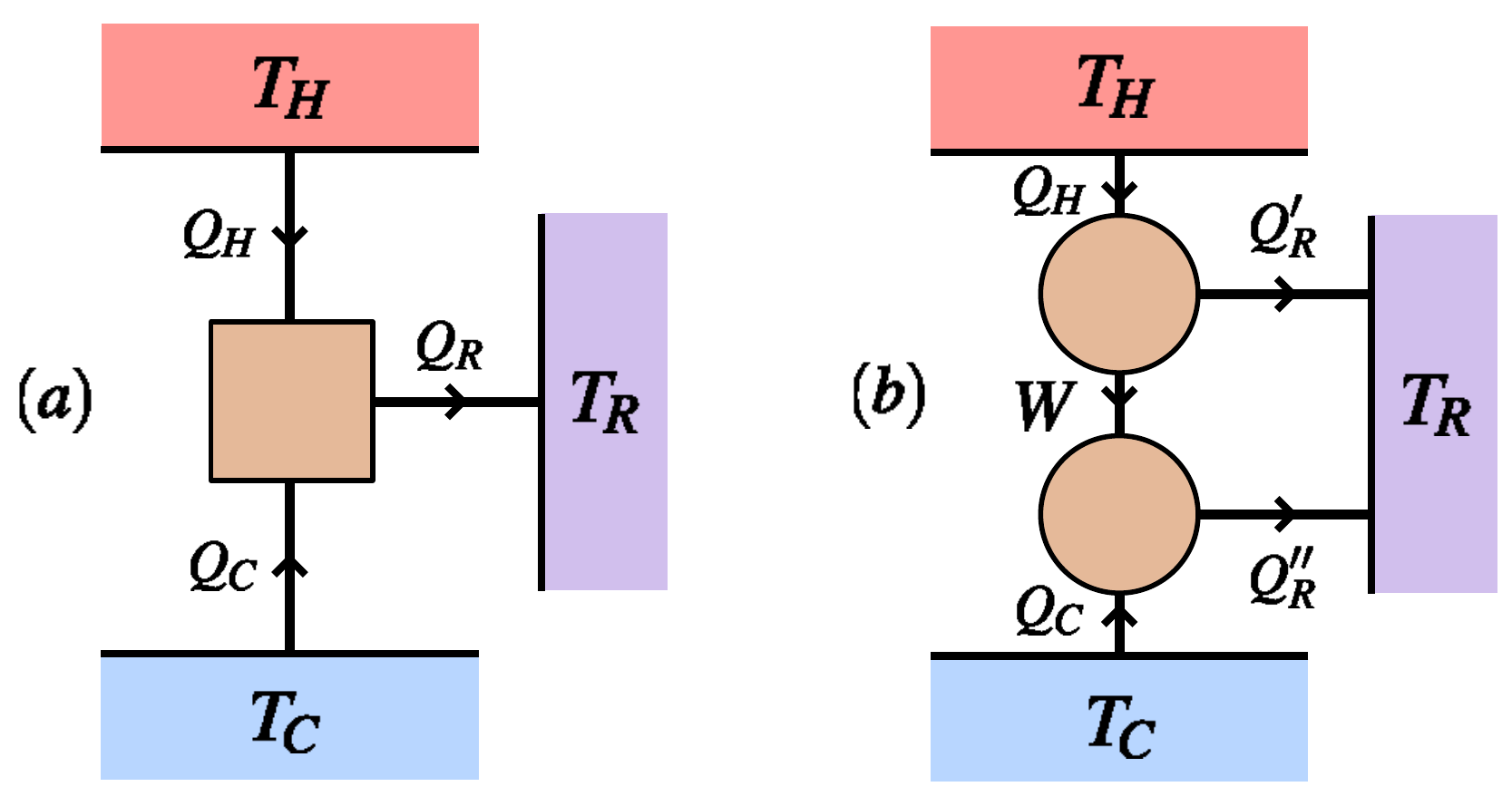} \caption{\label{f:engines}(a) Diagrammatic representation of a thermal machine which uses a supply of heat $Q_H$ extracted from a reservoir at $T_H$ to extract an amount of heat $Q_C$ from a reservoir at $T_C$, i.e. a refrigerator whose source of work is supplied by a thermal bath. (b) An explicit construction of such a device composed of two Carnot machines -- the top functioning as a heat engine, the bottom as a heat pump.} 
\end{figure}

In the standard analysis of the efficiency of a heat engine or refrigerator the efficiency is defined in terms of the \emph{work}; we are interested in how much work can be extracted from a given amount of heat, or how much heat can be extracted for a given amount of work. However, in the current scenario we have avoided the explicit notion of work -- the only free energy allowed is in the form of two baths at differing temperatures. We must therefore analyse the efficiency of such a device. Diagrammatically the machine that needs to be considered is depicted in Fig.~\ref{f:engines} (a).

That is, by extracting heat $Q_H$ from a hot reservoir at temperature $T_H$, we are able to extract an amount of heat $Q_C$ from a cold reservoir at temperature $T_C$ whilst `dumping' an amount of heat $Q_R$ into a reservoir at some intermediate temperature $T_R$. It follows that the appropriate measure of efficiency for such a machine is given by
\begin{equation}
	\eta = \frac{Q_C}{Q_H}
\end{equation}
that is, for a given supply of heat from a hot bath, how much heat can possibly be extracted from the cold bath. The two important points to note are first that the most efficient such machine will be a \emph{reversible} machine, just as in the case of all thermodynamic machines. The second point to note is that all reversible machines -- however they are constructed -- must run at the same efficiency, and therefore we can focus on a specific reversible model without loss of generality. The model we will focus on is comprised of a Carnot heat engine supplying an amount of work $W$ into a Carnot heat pump, as depicted in Fig.~\ref{f:engines} (b).

To calculate the efficiency of this machine, we first apply the first law of thermodynamics to the heat engine and heat pump separately to obtain
\begin{equation}\label{e:1st law}
	Q_H = Q_R^{\prime} + W,\quad\quad\quad\quad Q_C + W = Q_R^{\prime\prime},
\end{equation}
followed by the second law, telling us that entropy is conserved in a Carnot machine,
\begin{equation}\label{e:2nd law}
	\frac{Q_H}{T_H} = \frac{Q_R'}{T_R},\quad\quad\quad\quad\quad\quad \frac{Q_C}{T_C} = \frac{Q_R''}{T_R}.
\end{equation}
Equations \Eref{e:1st law} together imply that $Q_R'+Q_R^{''} = Q_H + Q_C$, which, when combined with \Eref{e:2nd law} leads to the Carnot efficiency for this machine,
\begin{equation}\label{e:carnot efficiency}
	\eta^{c} = \frac{Q_C}{Q_H} = \frac{1-\frac{T_R}{T_H}}{\frac{T_R}{T_C}-1}
\end{equation}
and is thus an upper bound on the efficiency of any such engine which we run between three reservoirs and which extracts heat from the bath at $T_C$ using a supply of heat from the reservoir at $T_H$. Note that when $T_R = T_H$ then the efficiency is zero; in this case we are unable to extract any work with the heat engine and thus are unable to power the heat pump. Conversely, when $T_R \rightarrow T_C$ it can be seen that $\eta^c$ diverges; in this limit heat can effectively be moved between the two reservoirs for `free'.

\section*{Conclusions}
By comparing the maximum quantum efficiency of our model \Eref{e:quantum max} and the Carnot efficiency, \Eref{e:carnot efficiency} we see a remarkable result: they coincide. Therefore the smallest possible refrigerator, despite the fact that it features a discrete and very small number of states, which one could have assumed to lead to stringent limitations on its efficiency, can actually achieve the maximum efficiency compatible with the laws of thermodynamics. In a further study \cite{virtual} we will argue that any thermal machine approaching the Carnot limit functions, essentially, as the smallest possible thermal machine.

\section*{Acknowledgments} We acknowledge support from EU integrated project Q-ESSENCE and the UK EPSRC.

\end{document}